\documentclass[12pt]{article}

\usepackage[margin=1.25in]{geometry}	
\usepackage{setspace}
\usepackage{graphicx}
\usepackage{hyperref}
\usepackage{url}
\usepackage{etoolbox}
\usepackage{bm}
\usepackage{amsmath,amssymb,amsthm}
\usepackage[linesnumbered,ruled,vlined]{algorithm2e}
\usepackage{float}
\usepackage{multirow}
\usepackage{colortbl}
\usepackage{enumerate}
\usepackage[table,xcdraw]{xcolor}
\usepackage{titlesec}
\titlelabel{\thetitle.\quad}
\usepackage{cite}

\begin{document}

\title{On Racial Disparities \\ in Recent Fatal Police Shootings}
\author{Lucas Mentch \vspace{7mm} \\
Department of Statistics \\ University of Pittsburgh\\ 
email: \texttt{lkm31@pitt.edu} \\ \hspace{15.5mm} }

\maketitle

\begin{center}
\vspace{15mm}
\textbf{Abstract}
\end{center}
Fatal police shootings in the United States continue to be a polarizing social and political issue.  Clear disagreement between racial proportions of victims and nationwide racial demographics together with graphic video footage has created fertile ground for controversy.  However, simple population level summary statistics fail to take into account fundamental local characteristics such as county-level racial demography, local arrest demography, and law enforcement density.  Utilizing data on fatal police shootings between January 2015 and July 2016, we implement a number of straightforward resampling procedures designed to carefully examine how unlikely the victim totals from each race are with respect to these local population characteristics if no racial bias were present in the decision to shoot by police.  We present several approaches considering the shooting locations both as fixed and also as a random sample.  In both cases, we find overwhelming evidence of a racial disparity in shooting victims with respect to local population demographics but substantially less disparity after accounting for local arrest demographics.  We conclude our analyses by examining the effect of police-worn body cameras and find no evidence that the presence of such cameras impacts the racial distribution of victims.

%%%%%%%%%%%%%%%%%%%%%%%%%%%%%%%%%%%%%%%%%%%%%%%%%%%%%%%%%%
%%%%%%%%%%%%%%%%%%%%%%%%%  INTRODUCTION  %%%%%%%%%%%%%%%%%%%%%%%
%%%%%%%%%%%%%%%%%%%%%%%%%%%%%%%%%%%%%%%%%%%%%%%%%%%%%%%%%%
\newpage
\doublespace
\section{Introduction}
The extensive media coverage of fatal police shootings in recent years in the United States has fueled political debate and sparked widespread controversy.  Due in part to this increased attention as well as concerns regarding federal data collection methods \cite{Kobler1975,Fyfe2002,Klinger2012,Nix2017,Klinger2017,Williams2019,Cesario2019,Klinger2016,White2016}, The Washington Post began compiling data on each fatal police shooting taking place in the U.S. beginning in 2015 \cite{WaPo}.  By raw totals, White victims far outweigh all other racial groups, accounting for nearly half (733) of the 1505 documented shootings between January 1, 2015 and July 11, 2016 \footnote{The version of the Washington Post dataset used here was accessed on July 12, 2016; the most recent shooting recorded at that time was said to have occurred on July 11, 2016.  The updated database can be found at \url{https://github.com/washingtonpost/data-police-shootings}.}.  However, when we compare the proportions of fatal shootings to the population demographics in the United States \cite{USCensus}, we see that the proportion of fatal shootings of Blacks is substantially higher than the population proportion, whereas the proportions of White and Asian fatal shooting victims fall below their respective population proportions; see Figure 1.  

If key population characteristics such as racial demography and law enforcement density could be assumed to be relatively uniform throughout the United States, this information alone could be considered sufficient to reasonably conclude that the racial proportions (and totals) of fatal police shootings are different from what would be expected under the assumption that race is independent of an officers decision to take potentially lethal action with a firearm.  This assumption, however, is simply not reasonable for a large, diverse area like the United States.  Thus, given the number of fatal police shootings that occurred between January 1, 2015 and July 11, 2016, the key questions we seek to address in the remainder of this paper are: \emph{Taking into account local characteristics, how many individuals from each race would be reasonable to expect if the fatal shooting victims could be seen as a random sample from the localized population and are the observed victim totals in line with such expectations.}  

In attempting to answer these, we make use of the fatal police shootings dataset compiled by the Washington Post as well as datasets containing county-level racial demography, law enforcement density, and local arrest demography.  Importantly, we stress that we examine the database of fatal police shootings in totality.  In particular, we make no attempt to segment these shootings into those which might be considered ``justified" or ``non-justified" and we do not consider whether or with what the victim may have been armed at the time of the shooting.  Though the dataset from the Washington Post does contain some information of this sort, it is difficult to determine in many instances whether a suspect ``armed" with, for example, a cell phone or a tape measure, actually attempted to present these items as weapons or whether they simply happened to be in their possession at the time of the incident. 

Studies related to police shootings and use of force have long produced a tremendous amount of literature; for a small sample of research from the past two decades, see for example \cite{Adams1999,Alpert2004,Alpert1997,Ariel2015,Fridell2016,Legewie2016,Smith2004,Lim2014,Paoline2007,Kop2001,Ridgeway2016,Fryer2016}.  As already eluded to, however, reliable data on this topic has proven extremely difficult to obtain with numerous studies continually finding underreporting in federal databases by as much as 50\% \cite{Kobler1975,Fyfe2002,Klinger2012,Nix2017,Klinger2017,Williams2019,Cesario2019,Klinger2016,White2016}.  Indeed, in line with this already well-established finding, the data from the Washington Post utilized here contains information on 515 fatal police shootings through July of 2016 whereas the FBI's Supplemental Homicide Report \cite{FBISHR} contains only 439 incidents for the entire year.  As remarked by Fyfe (2002) \cite{Fyfe2002} and later recalled by Klinger and Slocum (2017) \cite{Klinger2017}, it remains the case that ``the best data on police use of force come to us not from the government or from scholars, but from the Washington Post."  In light of this, researchers have recently begun focusing on more complete data provided by large journalistic outlets.  As one example, Nix et al.\ (2017) \cite{Nix2017} utilized the Washington Post data from 2015 to investigate incidents they determined to be ``threat-perception failures" and concluded that certain minority groups were less likely to be attacking an officer and/or armed at the time of the shooting.  Klinger and Slocum (2017) \cite{Klinger2017} take issue with this study however for reasons much in line with those noted in the preceding paragraph.  The authors argue that even unarmed individuals can pose a potentially serious threat and point to at least four separate incidents in which officers were attacked with objects that might be otherwise innocuous (e.g.\ metal pole, tree branch) and yet victims were categorized as ``unarmed" in the data provided by the Washington Post.  We emphasize that the work referenced above merely scratches the surface of all research on police-involved shootings.  For a more thorough accounting of existing research in this area, we refer the interested reader to the literature reviews provided in Ridgeway (2016) \cite{Ridgeway2016} and Nix et al.\ (2017) \cite{Nix2017} as excellent potential starting places.

Perhaps the study most similar in spirit to the work presented here was published very recently by Cesario et al.\ (2019) \cite{Cesario2019}.  Here too the authors point out the potential issues with seeking to identify bias by comparing the racial proportions of police shooting victims to nationwide racial demographics.  The authors instead argue that police are more likely to use deadly force in crime-related interactions and therefore utilize federal crime data to estimate national rates of criminal involvement for both Blacks and Whites.  Using police shooting data collected by The Guardian, they then compute the odds of both races being shot, ultimately concluding that no racial disparity exists relative to the estimated rates of criminal involvement. 

As noted above, the work here pushes beyond simple comparisons of nationwide proportions.  Instead, using the data collected by the Washington Post, we focus on local characteristics of the populations where police shootings actually took place in 2015 and 2016.  Furthermore, we utilize a resampling approach that allows us to estimate the entire distribution of the number of expected victims from each race under various setups rather than obtaining only a single number summary.   This approach thus allows us to more fully characterize the likelihood of observing the various counts actually observed during those years.  

The remainder of this work is laid out as follows.  A brief overview of the datasets is provided in Section 2 with a more thorough description and accounting given in the Appendix.  In Section 3 we employ a resampling scheme to estimate the distributions of total fatal police shooting victims by race, conditional on the locations where the observed shootings took place.  In Section 4 we consider an alternative scheme wherein the locations are selected at random and weighted according to relative law enforcement density and in Section 5 we incorporate local arrest demography into the analysis.  In Section 6 we compare the racial proportions of police shooting victims in incidents where the responding officers were wearing body cameras to those in which no body camera was present.  Finally, we conclude with a careful discussion of these results in Section 7.  In addition to the details provided in the Appendix, an accompanying \verb!R! file is also provided to reproduce all results and calculations.  

\begin{figure}
  \centering
  \includegraphics[scale=0.55]{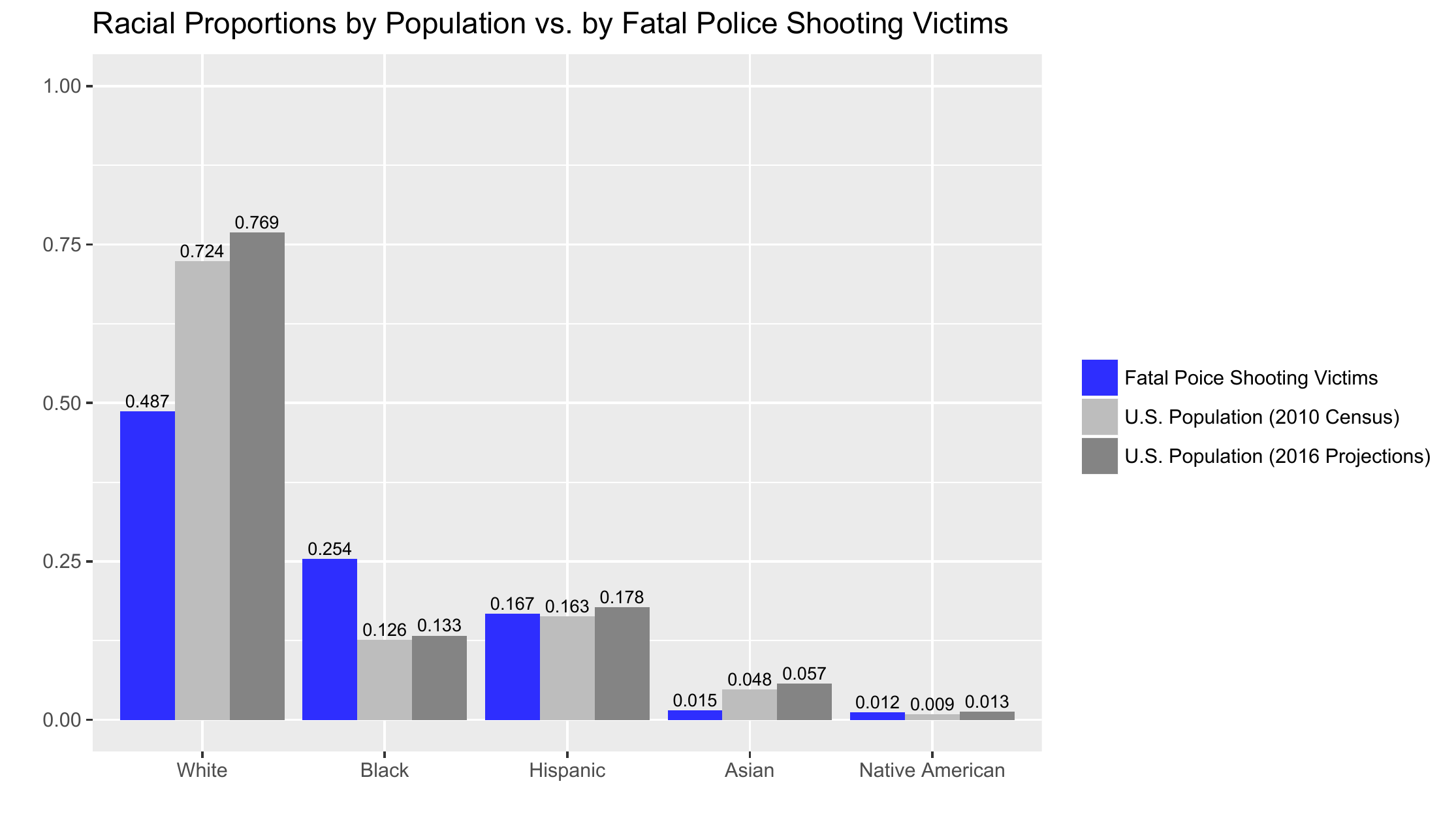}
  \vspace{-7mm}
  \caption{\label{fig:barchart1} Racial proportions by total U.S. population vs. by fatal shooting victims.  Fatal shooting proportions were tabulated directly from the data provided by The Washington Post \cite{WaPo} dataset; population proportions are shown according to the U.S. Census Bureau \cite{USCensus}.}
\end{figure}

%%%%%%%%%%%%%%%%%%%%%%%%%%%%%%%%%%%%%%%%%%%%%%%%%%
%%% Data Overview
%%%%%%%%%%%%%%%%%%%%%%%%%%%%%%%%%%%%%%%%%%%%%%%%%%
\section{Data Overview}

The analyses in the following sections make use of a total of five different publicly accessible datasets.  Here we provide a short overview of each.  Appendix B contains more detailed information including access instructions as well as a thorough accounting of modifications and corrections made to the original data in order to perform the analyses.  The abbreviations and numeric citations listed indicate how each individual dataset will be referenced in future sections.

\begin{itemize}
\item {\bf WP \cite{WaPo}:  }  The primary dataset of interest here, containing information on recent fatal police shootings as collected and reported by the Washington Post.  Note that this dataset contains only instances of \emph{fatal shootings}; nonfatal shootings and other police encounters resulting in death are not included.  In addition to the date, city, and state of these fatal shootings, the dataset also contains a number of other features such as what (if anything) the victim was armed with, an indicator for whether the responding law enforcement officers were wearing body cameras, an indicator for whether the victim displayed signs of mental illness, the victim's (estimated) threat level, in what fashion (if at all) the victim was fleeing, as well as age, gender, and race of the victim \footnote{Note that as can be inferred from Figure \ref{fig:barchart1}, the WP dataset treats `Hispanic' as a possible race rather than as an ethnicity that crosses racial categories, as defined in the census data.}.  Based on the city and state information, county information was later imputed according to information provided by the National Association of Counties (NACo) \cite{NACo}.  

\item {\bf DEM \cite{USCBdem}:  }  This dataset contains information on county-level racial demographics in the United States based on the 2010 U.S. Census.  Information is provided for a total of 3142 counties or parishes.  For each county or parish, the total population is provided along with the total population of a particular race.  The races included are:  White (W), Black (B), Native American and/or Alaskan Native (NA), Asian (A), Native Hawaiian and Other Pacific Islander (NH), and finally ``Two or More'' (T).  While it's generally easier to refer to this dataset as a single object, in each of the resampling analyses, we make explicit use of two separate datasets containing this demographic information:  one based on the 2010 census data itself and another based on the projected demographics in 2016.

\item {\bf LEE \cite{FBItable80}:  }  This dataset contains information on county-level \textbf{L}aw \textbf{E}nforcement \textbf{E}mployment collected by the FBI through the 2011 Uniform Crime Reporting (UCR) Program.  Information is provided for a total of 2797 counties or parishes.  For each county or parish, the total number of law enforcement employees is provided and broken down by officers and civilians.  From the FBI data disclosure, ``the UCR Program defines law enforcement officers as individuals who ordinarily carry a firearm and a badge, have full arrest powers, and are paid from governmental funds set aside specifically to pay sworn law enforcement'' whereas ``civilian employees include full-time agency personnel such as clerks, radio dispatchers, meter attendants, stenographers, jailers, correctional officers, and mechanics''.

\item {\bf ARREST \cite{ICPSR36115}:  }  This dataset contains information on local, county-level arrests by age, sex, and race in 2013.  For each reporting county, a number of different offenses are reported and for each offense type, the total number of arrests made according to age, sex, and race demographics is provided.  In total, 2754 counties report at least one crime.  The races included in this dataset are White (W), Black (B), Native American and/or Alaskan Native (NA), and Asian (A).  County information is provided by FBI UCR numeric code instead of name. 

\item {\bf CODES \cite{ICPSR2565}:  }  This dataset contains state, county, and parish names along with Uniform Crime Reports (UCR) and Federal Information Processing Standards (FIPS) numeric codes.  The primary purpose of this dataset is to facilitate linking between the ARREST data which contains only UCR county codes and the WP, DEM, and LEE datasets which contain only the county or parish names.

\end{itemize}

%%%%%%%%%%%%%%%%%%%%%%%%%%%%%%%%%%%%%%%%%%%%%%%%%%%%%%%%%%%%
% Local Population Demographics
%%%%%%%%%%%%%%%%%%%%%%%%%%%%%%%%%%%%%%%%%%%%%%%%%%%%%%%%%%%%
\section{Local Population Demographics}
We now begin the resampling analyses to investigate how unlikely the observed racial distributions of fatal police shooting victims between January 2015 and July 2016 would be if the victims of these shootings could be considered a random sample from the local population in which the shootings took place and no racial biases were present.  As discussed above, it can be readily seen from Figure 1 that the racial proportions of victims appear out-of-line with the \emph{nationwide} racial demographics; here we utilize the WP and DEM datasets to determine whether the same can be said after taking into account the \emph{local} racial demographics.  

Of the 1505 fatal shootings in the WP dataset, only one shooting location did not appear in the DEM dataset; this was WP ID number 686 which occurred in  Las Cruces, NM in Do\~{n}a Ana County.  Furthermore, there are a total of 77 additional victims in the WP dataset for which race information is missing.  After removing these, there are a total of 1427 victims for which both the race of the victim and the county-level racial demographics are known.  Of these, 733 (51.4\%) were White, 382 (26.8\%) were Black, 251 (17.6\%) were Hispanic, 18 (1.3\%) were Native American, 22 (1.5\%) were Asian, and 21 (1.5\%) are listed as `Other'.  

To investigate the plausibility of this observed racial distribution, we performed 1000 simulations in which an individual race was selected at random from each shooting location according to the racial proportions in the county in which the shooting occurred.  That is, if a particular shooting occurred in a county in which the proportions were 50\% White, 20\% Black, 5\% Native American, and 5\% Asian, then W, B, N, or A would be selected with probabilities 0.5, 0.2, 0.05, and 0.05, respectively.  The total number of victims from each race were then summed and the entire process repeated 1000 times.  Thus, for each race, we obtain an estimated distribution of victim totals under the assumption that fatal shooting victims can be considered a random sample from the racial demographics of the county in which the shooting occurred.  

Before undertaking these simulations, we first need to address the disagreement in racial categories between the WP and DEM datasets:  the WP dataset contains the racial categories W, B, NA, A, H, and O while the DEM datasets contain the categories W, B, NA, A, NH, and T.  Beyond these racial categories, the DEM datasets also contain information on how many residents of each race are Hispanic.  Given this additional information, accounting for the Hispanic (H) population in our resampling procedure is straightforward:  once a race is selected, we either keep that race or replace it by `Hispanic' with probability weighted according to the local Hispanic population.  More formally, let $H_{ij}$ denote the proportion of race $i$ that is Hispanic in county $j$.  Supposing that race $i$ is that which is first randomly selected, we record a victim belonging to that race with probability $1-H_{ij}$ and instead record a race of `Hispanic' with probability $H_{ij}$.

We now finally need to determine whether the O classification in the WP dataset roughly corresponds to the combined NH and T classifications in the DEM datasets.  Looking at these datasets, we see that the average proportion of the population classified as either NH or T is approximately 2.48\% according to the 2010 census and 2.86\% according to the 2016 projections.  According to the U.S.\ Census Bureau Quickfacts \cite{USCensus}, in 2016, W, B, NA, and A made up approximately 76.9\%, 13.3\%, 1.3\%, and 5.7\% of the population respectively, leaving 2.8\% of the population for an `Other' category.  Since this is in close agreement with the raw averages from the DEM datasets, we proceed accordingly treating NH and T in the DEM datasets as the equivalent of the O classification in the WP dataset.  That is, whenever either NH or T is selected in the resampling procedure, we first randomly determine whether to count the observation as Hispanic and if not, we count the observation as Other (O).

\begin{figure}
  \centering
  \includegraphics[scale=0.6]{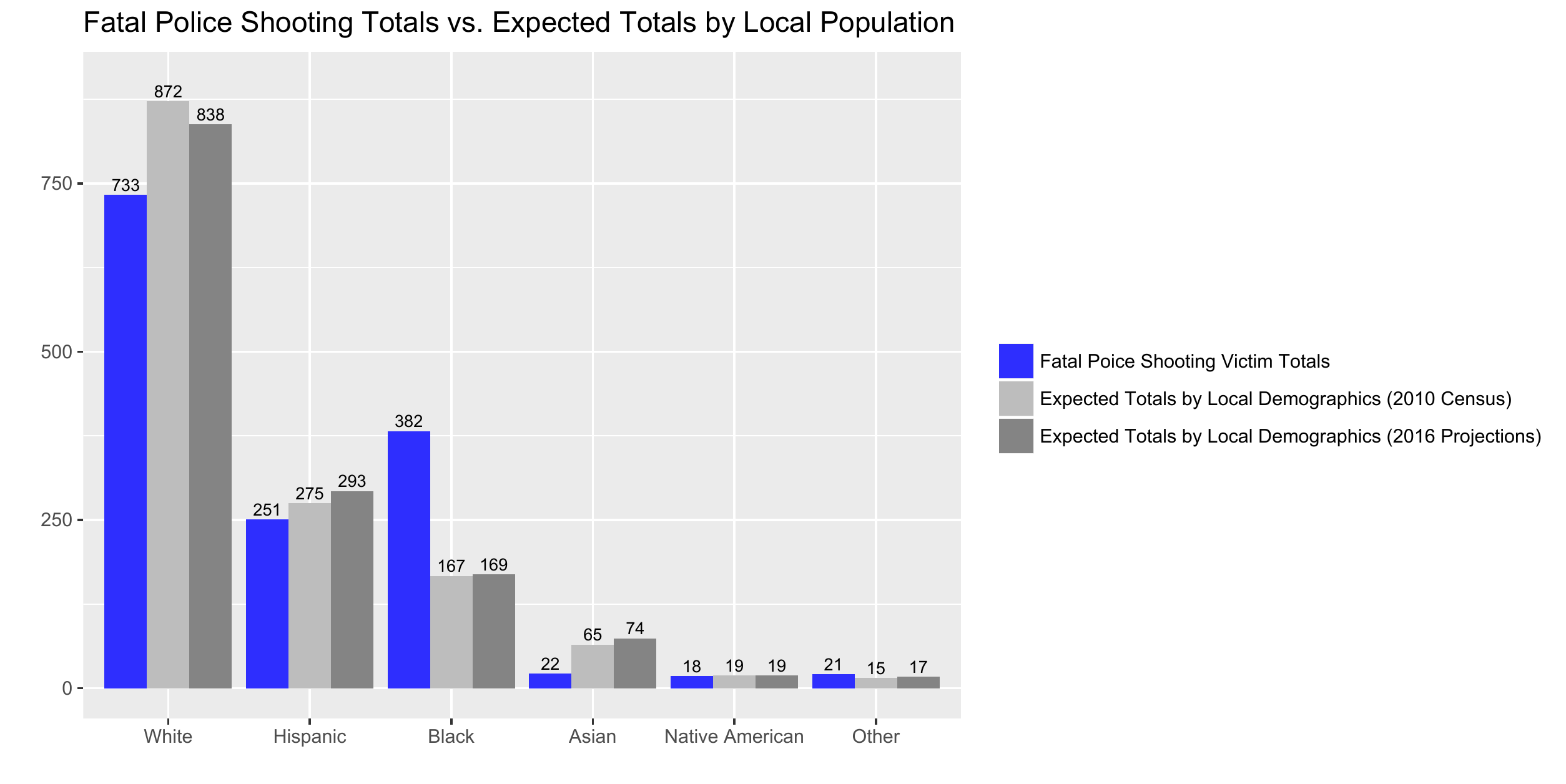}
  \vspace{-7mm}
  \caption{\label{fig:barchart2} Fatal police shooting victim totals vs.\ expected totals by local racial demographics assuming a fixed shooting location.}
\end{figure}

Figure \ref{fig:barchart2} shows the total number of police shooting victims by race for each of the six classifications given in the WP dataset along with the expected totals with respect to both the 2010 census data and the 2016 projections.  These expected totals for each race were computed by simply taking the mean of the 1000 resampled victim counts and rounding to the nearest integer.  Here we see the same general pattern as in Figure 1:  while the total number of Native American victims seems to be in line with what would be expected, the observed totals are fewer than expected for Whites and Asians while much greater than expected for Blacks.  In fact, according to these results, more than twice as many Blacks were fatally shot by police than would be expected if the shooting victims could be considered a random sample from the local population.  Note that Figure \ref{fig:barchart2} describes only how the true victim totals compare to the mean of the observed empirical distribution; density estimates for total victims of each race are shown in Figure 3 in Section 4. 

In order to formalize these results, we can obtain p-value estimates to assess whether the observed victim totals are significantly different from what would be expected by comparing the observed total to the totals found in our resampling procedure and counting the number of resampled totals that were more extreme than that observed.  The probability of observing a victim total more extreme than that observed is calculated as
\[
\hat{p}_r = 2 \times \frac{\min \{ \# \text{greater than } T_r, \, \# \text{less than } T_r \} }{N}
\] 

\noindent where $T_r$ denotes the observed victim total for race $r$ so that $\#$ greater (or less) than $T_r$ counts the number of estimates of victim totals out of the $N=1000$ that resulted in a victim count more (or less) than the observed total.  These estimates take the form of binomial random variables and thus have standard deviation $\sqrt{p_r (1-p_r) / N}$.  We don't know the true (exact) p-value $p_r$ but the standard deviation is bounded above by $\frac{1}{2 \sqrt{N}} \approx 0.015$.  For a further discussion on these errors we refer the reader to \cite{Efron1994,Good2005,Ojala2010}.

A table of such p-values is shown in Table 1.  Note that p-values are all 0 for Whites, Blacks, and Asians with respect to both the 2010 census and 2016 projected racial demographics indicating that the observed victim totals are significantly different from what would be expected.  On the other hand, we see highly non-significant results for Native Americans while for Hispanics and Other, we see significant results at an $\alpha$-level of 0.05 based on the 2016 projected racial demographics, but not according to the 2010 census.  Results remaining significant after a Bonferroni correction are shown in bold.

\noindent \textbf{\emph{Remark:  }}  The p-value formula above of the form $m/N$ represents the standard unbiased estimate where $N$ denotes the total number of resamples and $m$ denotes the number of resulting statistics more extreme than that originally observed.  In permutation/randomization-test settings, some (see \cite{Phipson2010} for example) have instead advocated for a  biased estimate of the form $(m+1)/(N+1)$ that accounts for the original statistic in order to ensure that the estimated p-value does not inflate the type I error rate of the resulting test.  All tests performed in this paper involve a large number of resamples ($N \geq 1000$) and thus the two estimates are nearly equivalent, but nonetheless, estimates of the latter form can be easily calculated from the tables of p-values given throughout the remainder of this paper.

%%%%%%%%%%%%%%%%%%%%%%%%%%%%%%%%%%%%%%%%%%%%%%%%%%
%%% Random shooting locations 
%%%%%%%%%%%%%%%%%%%%%%%%%%%%%%%%%%%%%%%%%%%%%%%%%%
\section{Treating the Shooting Locations as Random}

The preceding analysis suggests strongly that the racial proportions of fatal police shooting victims are not representative of the racial demography of the counties in which those shootings occurred.  It may be reasonable, however, to consider that there is some randomness involved with the shooting locations themselves.  For example, if police are led on a high speed chase across counties that ends with the suspect dead after a shootout, this would count as a fatal police shooting that occurred in that final county and would ignore information about the county in which the incident originated.  As another example, it is entirely possible that around the same time when some of these incidents occurred, other unrelated \emph{nonfatal} shootings occurred under similar circumstances elsewhere in the U.S.  Since the WP dataset contains only information on \emph{fatal} police shootings, these other possible incidents remain unaccounted for in the previous analysis.

Thus, we now consider a setup whereby the locations of the 1427 shootings are taken as a random sample from all U.S. counties, weighted by law enforcement officer employment.  That is, instead of choosing a race according to the local racial demographics of the counties in which these shootings actually occurred, we instead select counties at random with those employing a larger number of law enforcement officers being more likely to be selected.  

In order to perform this kind of resampling, we make use of the LEE dataset described in Section 2 that contains law enforcement employment totals for both officers and civilians.  In our procedure, we resample according to law enforcement \emph{officer} employment as these individuals are specifically defined as those who ``ordinarily carry a firearm and a badge'' \cite{FBItable80}.  Note that while we could instead weight the resampling by \emph{total} law enforcement employment, we do not expect that this alternative approach would produce significantly different results as the correlation between officer employment and total employment is exceptionally high at 0.97.

\begin{figure}
  \centering
  \includegraphics[scale=0.6]{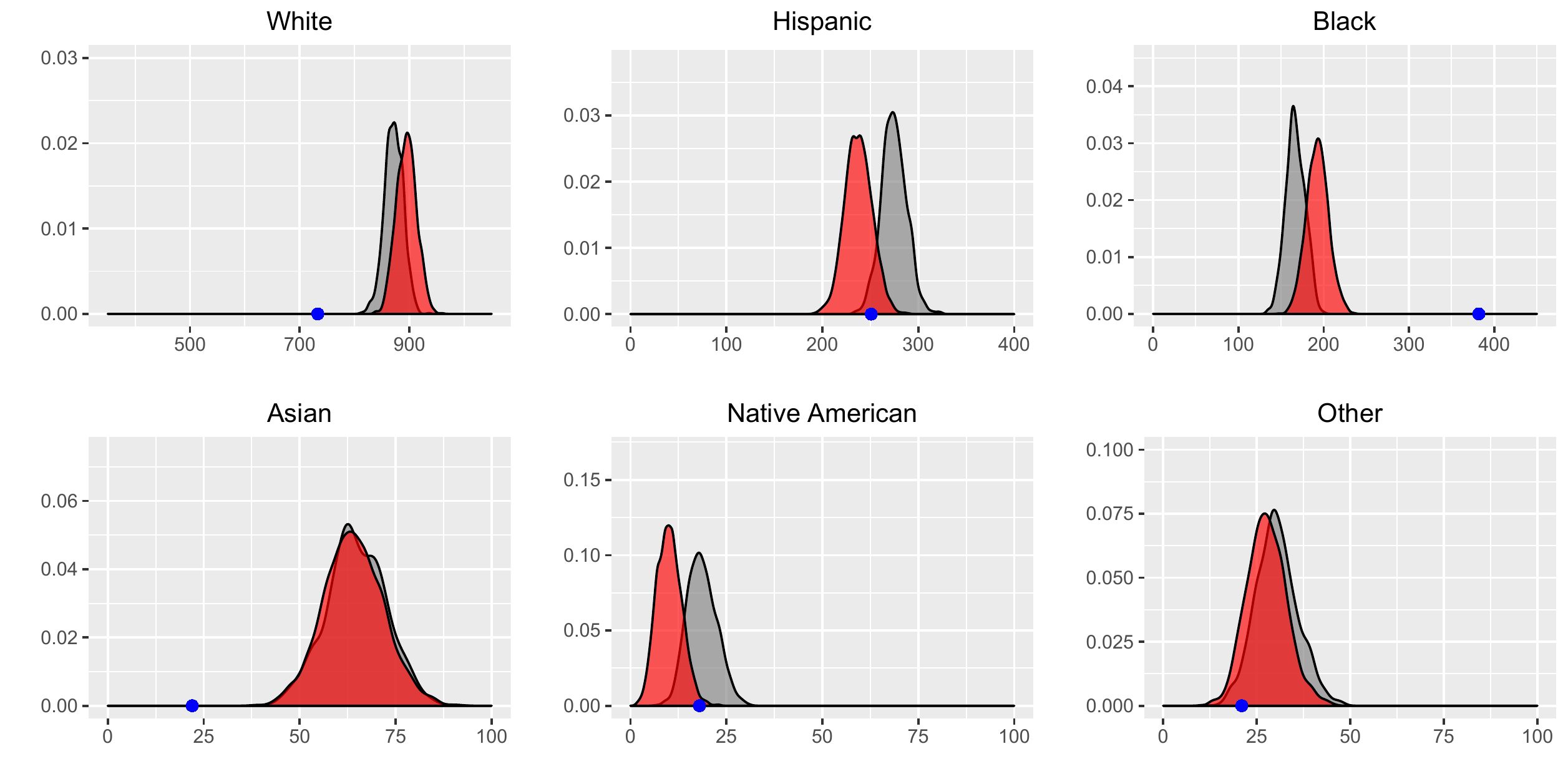}
  \vspace{-7mm}
  \caption{\label{fig:density1} Kernel density estimates for victim totals by race assuming fixed shooting locations (gray) or random locations (red) with respect to the 2010 census data on local racial demographics.  Blue points along each horizontal axis correspond to the observed victim totals in the WP dataset.}
\end{figure}

The remainder of the resampling procedure is identical to that laid out in Section 3, except that this time we employ a total of 2000 simulations -- double the previous number -- in order to account for the additional randomness involved in selecting a location.  The same reasoning as the previous section gives that the standard deviation of the resulting p-value estimates should be bounded above by $\frac{1}{2 \sqrt{2000}} \approx 0.011$.  The empirical densities of victim totals by race with respect to the 2010 census data are shown in Figure \ref{fig:density1}.  The densities with respect to the 2016 projections are extremely similar and the corresponding plots are thus reserved for Figure \ref{fig:density2016} in Appendix C.  Note that for the Asian and Other races, the densities of victim totals based on fixed locations are nearly identical to those where locations were selected at random.  For the White and Black races, the densities based on random locations appear shifted slightly right (higher expected totals) while for the Hispanic and Native American races, the densities based on random locations are shifted more substantially left (lower expected totals).  

Though the densities appear to shift a bit for some races when considering the shooting locations as random, overall the same general patterns appear to be present:  for the Hispanic, Native American, and Other races, the observed victim totals (blue points in Figure \ref{fig:density1}) appear more reasonable based on the empirical distributions while for the White, Black, and Asian races, the observed totals lie far from the densities estimated via the resampling procedures.  To examine this more formally, we can compute p-values in the same fashion as described in Section 3 with respect to these new densities based on random shooting locations.  A table of these p-values along with those calculated assuming a fixed shooting location is shown in Table \ref{tab:pvals}.  

\begin{table}[]
\centering
\begin{tabular}{|r|c|c|c|c|}
\hline
\multicolumn{1}{|l|}{}   & \multicolumn{2}{c|}{\textbf{Fixed Locations}}             & \multicolumn{2}{c|}{\textbf{Random Locations}}        \\ \hline
\multicolumn{1}{|l|}{}   & \textbf{2010 Census}      & \textbf{2016 Proj.}     & \textbf{2010 Census}      & \textbf{2016 Proj.} \\ \hline
\textbf{White}           & \cellcolor[HTML]{C0C0C0}$\bm{0}$ & \cellcolor[HTML]{C0C0C0}$\bm{0}$     & \cellcolor[HTML]{C0C0C0} $\bm{0}$ & \cellcolor[HTML]{C0C0C0}$\bm{0}$ \\ \hline
\textbf{Hispanic}        & 0.078                     & \cellcolor[HTML]{C0C0C0}$\bm{0.002}$ & 0.333                     & 0.836                     \\ \hline
\textbf{Black}           & \cellcolor[HTML]{C0C0C0}$\bm{0}$ & \cellcolor[HTML]{C0C0C0}$\bm{0}$     & \cellcolor[HTML]{C0C0C0}$\bm{0}$ & \cellcolor[HTML]{C0C0C0}$\bm{0}$ \\ \hline
\textbf{Asian}           & \cellcolor[HTML]{C0C0C0}$\bm{0}$ & \cellcolor[HTML]{C0C0C0}$\bm{0}$     & \cellcolor[HTML]{C0C0C0}$\bm{0}$ & \cellcolor[HTML]{C0C0C0}$\bm{0}$ \\ \hline
\textbf{Native Am.} & 0.97                      & 0.984                         & \cellcolor[HTML]{C0C0C0}0.018                     & \cellcolor[HTML]{C0C0C0}0.024                     \\ \hline
\textbf{Other}           & 0.102                     & \cellcolor[HTML]{C0C0C0}0.02  & 0.239                     & 0.056                     \\ \hline
\end{tabular}
\caption{\label{tab:pvals}P-values based on the empirical distribution of victim totals from each race based on the 2010 census data and 2016 projections.  Gray cells correspond to values significant at level $\alpha=0.05$. Results remaining significant after a Bonferroni correction are shown in bold.}
\end{table}

Looking at Table \ref{tab:pvals}, it is immediately clear that taking into account local population demographics does not help explain the victim totals observed for Whites, Blacks, and Asians.  Assuming a level of $\alpha = 0.05$,  for Hispanics and Others, we would reject the null hypothesis that the observed victim totals could have come from the distributions based on the 2016 projections with fixed shooting locations, while we fail to reject in every other case.  For Native Americans, we fail to reject the null hypothesis for the densities based on fixed shooting locations, but would reject it according to the densities estimated assuming random shooting locations.

%%%%%%%%%%%%%%%%%%%%%%%%%%%%%%%%%%%%%%%%%%%%%%%%%%%%%%%%%%%%
% Crime Demographics
%%%%%%%%%%%%%%%%%%%%%%%%%%%%%%%%%%%%%%%%%%%%%%%%%%%%%%%%%%%%
\section{Local Arrest Demographics}
\label{sec:crime}
In the previous analyses, we investigated how many victims from each race might be expected if the victims were thought of as a random sample from the local population.  For some races -- White, Black, and Asian, in particular -- the expected victim totals were far from what was observed in the WP dataset, suggesting that this assumption is a bit too na\"ive and unreasonable.  In very recent work, Cesario et al.\ (2019) \cite{Cesario2019} make a similar point, arguing that disparities in police shooting rates across race should be investigated relative to rates of criminal involvement.  The authors attempt to estimate nationwide crime rates for both Blacks and Whites, focusing in particular on murder/nonnegligent manslaughter, violent crime, and weapons violations. Such crimes, they claim  ``are the most aggressive in terms of interpersonal violence and, as such, are appropriate proxies for exposure to those situations during which police may be more likely to use deadly force" \cite{Cesario2019}.  The authors ultimately find no evidence of anti-Black disparities relative to their estimated crime rates.   

Here, rather than estimate a single nationwide rate of criminal involvement for each race, we make use of the ARREST data described in Section 2 which contains information on local, county-level arrests by offense, age, sex, and race.  Since county information is provided only by FBI UCR numeric codes, the CODES dataset was also used to impute county names to allow for cross-reference with the datasets used in the previous analyses.  The objective here is to employ the same sort of resampling scheme utilized in the previous analyses except that instead of sampling victim races by weighting with respect to local population demographics, here we weight samples according to local arrest demographics.

Complicating matters is the fact that the ARREST dataset consists of only four racial categories:  W, B, NA, and A.  If we were to sample from only these races, we would necessarily overestimate the victim totals for each race and thus we need to be sure to account for the additional racial categories -- Hispanic (H) and Other (O) -- appearing in the WP dataset.  Though somewhat unsatisfying, the most reasonable way to accommodate this given the data is to make the assumption that the proportion of arrests in each county of individuals belonging to these additional races are the same as the population proportions.  Thus, under this setup, we expect the distributions for H and O to be similar to what was seen in the previous analyses.  Fortunately however, as seen in those previous analyses, the Hispanic and Other races were two in which the number of observed victims seemed to be reasonable with respect to local population demographics.  For these reasons, we focus our attention here more heavily on those races -- W, B, NA, A -- for which we have the arrest data.  

It's also worth pointing out that to employ this sort of resampling procedure, we need for every location in the WP dataset to contain information in the ARREST dataset and also, because we need to impute information from the DEM dataset into the arrest dataset (to account for the H and O populations), we need entries in the ARREST dataset to have information in the DEM dataset.  Because of the disagreement in county information across these three datasets, we need to further subset the WP data to include a total of only 1249 fatal police shooting victims (W (654), B (314), NA (12), A (21), H (229), O (19)).  Further details are provided in the appendix and the accompanying \verb!R! file. 

We also need to consider how the population of arrestees should be sampled.  Given the breakdown by race for each type of offense, we could consider some sort of weighted approach whereby, for example, violent offenses have a higher chance of being selected in a similar spirit to Cesario et al.\ (2019) \cite{Cesario2019}.  Indeed, we could potentially assume that individuals previously arrested for violent crimes would be more likely to present a substantial threat to police officers and/or that officers may be more likely to perceive a threat from such known individuals.  However, this approach would necessarily involve not only making this additional assumption, but would also require a subjective judgement with regard to what crimes should be considered and how those crimes should be weighted.  The decision by Cesario et al.\ (2019) to focus on ``murder/nonnegligent manslaughter, violent crime, and weapons violations" appears to be largely based on personal belief rather than being grounded in hard evidence.  There have certainly been numerous highly-visible instances in recent years in which fatal police shootings have occurred following what began as relatively routine encounters.  While in theory such assumptions could be checked against a national database like the FBI's Supplemental Homicide Report \cite{FBISHR}, as discussed in the introduction, such records are notoriously incomplete and likely suffer from bias as a result of voluntary self-reporting.  Given the lack of strong empirical evidence suggesting that all or even the vast majority of police shooting victims are violent criminals, we elect to not make such assumptions and instead aggregate arrest totals across offense types creating a single row of data for each county corresponding to the total number of arrests per race in that county. 

\begin{figure}
  \centering
  \includegraphics[scale=0.6]{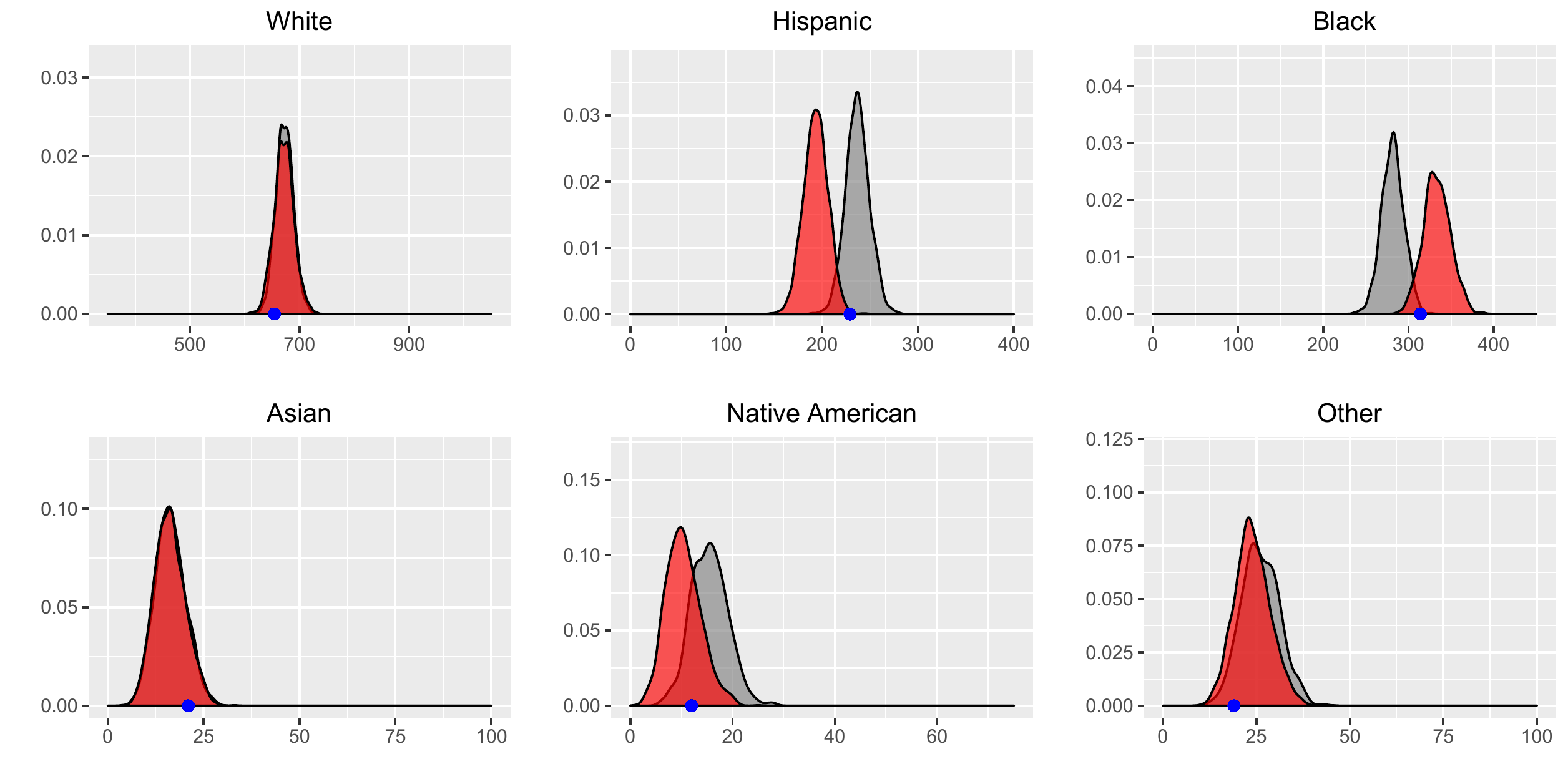}
  \vspace{-7mm}
  \caption{\label{fig:density3} Kernel density estimates for victim totals by race assuming fixed shooting locations (gray) or random locations (red).  Blue points along each horizontal axis correspond to the true (observed) victim totals from that race in the WP dataset.  Victim races selected according to local arrest demographics with races missing arrest totals (Hispanic and Other) sampled according to the 2010 census data.}
\end{figure}

As in the previous analyses, we consider resampling procedures based on the fixed locations in which the observed shootings took place and also consider the randomized approach whereby counties are selected according to police employment totals.  Because information needs to be imputed for the H and O populations, each procedure was also run twice:  once using the 2010 census data to perform the imputation and once using the 2016 demographic projections.  The resulting density estimates calculated with respect to the 2010 census data are shown in Figure \ref{fig:density3}; the densities calculated using the 2016 projections are very similar and thus are reserved for Figure \ref{fig:density4} in Appendix C.

It is immediately clear based on a quick visual inspection that the observed victim totals appear much more reasonable with respect to these distributions than with respect to those based only on local population demographics as investigated in Sections 3 and 4.  To be thorough, we compute p-values in the same fashion as in those previous sections with respect to these new densities.  A table of these p-values is shown in Table \ref{tab:pvalsCRIME}.  Note that Hispanic and Black are the only races for which a significant result is found at the $\alpha=0.05$ level and even for these races, the estimated totals are significant with respect to only two distributions and not significant with respect to the other two.  No results remain significant after a Bonferroni correction. 

Furthermore, examining these p-values alone can be somewhat misleading.  For both the Black and Hispanic races, we can see visually from the densities in Figures \ref{fig:density3} and \ref{fig:density4} that the observed victim totals seem to lie ``in between'' the densities based on fixed vs.\ random locations.  Thus, while the observed Hispanic victim total is somewhat higher than might be expected if the locations are assumed to be random, it is somewhat lower than what would be expected if the locations are seen as fixed.  For Black victims on the other hand, the opposite is true:  the number of victims observed is on the very high end of what would be expected assuming fixed locations, but on the low end of what would be expected if the shooting locations are seen as random.  In the way of final confirmation that these distributions are more appropriate, Table \ref{tab:SDdistances} in the Appendix shows the number of standard deviations the observed totals lie from the expected totals with respect to each distribution estimated in these and previous analyses.  For Whites, Asians, and Blacks, the observed totals fall at least 6, 5, and 14 standard deviations from the expected totals when the resampling is done with respect to local population demographics.  When the resampling is done instead with respect to local arrest demographics, the observed totals lie mostly within 1 to 2 standard deviations of what would be expected.

\begin{table}[]
\centering
\begin{tabular}{|r|c|c|c|c|}
\hline
\multicolumn{1}{|l|}{}   & \multicolumn{2}{c|}{\textbf{Fixed Locations}}                 & \multicolumn{2}{c|}{\textbf{Random Locations}}            \\ \hline
\multicolumn{1}{|l|}{}   & \textbf{2010 Census}          & \textbf{2016 Proj.}     & \textbf{2010 Census}          & \textbf{2016 Proj.} \\ \hline
\textbf{White}           & 0.266                         & 0.936                         & 0.333                         & 0.925                     \\ \hline
\textbf{Hispanic}        & 0.540                         & \cellcolor[HTML]{C0C0C0}0.046 & \cellcolor[HTML]{C0C0C0}0.003 & 0.116                     \\ \hline
\textbf{Black}           & \cellcolor[HTML]{C0C0C0}0.004 & \cellcolor[HTML]{C0C0C0}0.008 & 0.229                         & 0.287                     \\ \hline
\textbf{Asian}           & 0.220                         & 0.198                         & 0.209                         & 0.212                     \\ \hline
\textbf{Native Am.} & 0.412                         & 0.478                         & 0.509                         & 0.409                     \\ \hline
\textbf{Other}           & 0.186                         & 0.064                         & 0.334                         & 0.116                     \\ \hline
\end{tabular}
\caption{\label{tab:pvalsCRIME}P-values based on the empirical distribution of victim totals from each race based on the 2010 census data and 2016 projections.  Gray cells correspond to values significant at level $\alpha=0.05$. No results remain significant after a Bonferroni correction. }
\end{table}

%%%%%%%%%%%%%%%%%%%%%%%%%%%%%%%%%%%%%%%%%%%%%%%%%%%%%%%%%%%%
%  Body Camera Data
%%%%%%%%%%%%%%%%%%%%%%%%%%%%%%%%%%%%%%%%%%%%%%%%%%%%%%%%%%%%
\section{Body Camera Effects}

Finally, yet another interesting feature of the WP dataset is an indicator variable for whether the officers involved were wearing body cameras.  Recall that of the 1505 total fatal shootings recorded in the WP dataset, 1428 contain the (known) race of the victim.  Amongst these instances, the officers involved were wearing body cameras 132 times and thus no body cameras were present in the majority (1296) of instances.  The breakdown by race and body camera is shown in Table \ref{tab:chisq} as a standard $\chi^2$ contingency table.

We conclude our analyses by assessing whether there is a difference in the racial proportions of victims when body cameras were being worn by the officers involved.  A quick look at Table \ref{tab:chisq} reveals that the observed and expected cell counts appear to be very close and carrying out the $\chi^2$ test (degrees of freedom $(2-1)(6-1)=5$) confirms these suspicions (test statistic value of 5.17; $p$-value $= 0.395$).  According to this test and these data, there is no evidence that would suggest a significant difference in the racial proportions of victims whenever a body camera is in use.  It's worth noting however that the small expected counts for Native Americans, Asian Americans, and Others in the body camera group are something of a concern as these expected count values should generally be larger in order for the $\chi^2$ distribution to serve as a good approximation. 

\noindent \textbf{\emph{Remark:  }}  Here we are specifically interested in whether the distribution of victims across race remains the same when police-worn body cameras are vs.\ are not in use and thus a standard $\chi^2$ testing approach is most natural.  While in theory a similar approach could be used to evaluate the hypotheses in previous sections, this would result in an extremely large contingency table with $6 \times 2797 = 16,782$ cells -- one per race for each of the 2797 counties in the LEE dataset -- the vast majority of which would necessarily have a count of 0 given that we have only 1428 total shootings.  For a recent overview of the difficulties and various approaches to dealing with these and related high-dimensional testing issues, we refer the interested reader to \cite{Balakrishnan2018}.  For completeness, in addition to the $\chi^2$ test above, we also provide a randomization test similar to those in previous sections for assessing the effect of body cameras in Appendix D.  The randomization tests suggest a significant effect at the $0.05$ level only for Native Americans; no significant results remain after a Bonferroni correction.

\begin{table}[]
\footnotesize
\centering
%\hspace{-50mm}
\begin{tabular}{|r|c|c|c|c|c|c|c|}
\hline
%\multicolumn{1}{|c|}{} & W        & B        & \begin{tabular}[c]{@{}c@{}}Native \\ American\end{tabular} & Asian       & \multicolumn{1}{l|}
\multicolumn{1}{|c|}{} & W        & B        & NA & A  & H & O     & \multicolumn{1}{l|}{\textbf{Total}} \\ \hline
BodCam            & 64 (67.8)   & 38 (35.3)   & 4 (1.7)    & 1 (2.0) & 24 (23.3) & 1 (1.9)   & \textbf{132}                        \\ \hline
No BodCam         & 669 (665.2) & 344 (346.7) & 14 (16.3)  & 21 (20.0) &  228 (228.7) & 20 (19.1) & \textbf{1296}                       \\ \hline
\textbf{Total}         & \textbf{733} & \textbf{382} & \textbf{18}                                                & \textbf{22} & \textbf{252} & \textbf{21} & \textbf{1428}                       \\ \hline
\end{tabular}
\caption{\label{tab:chisq} Race vs Body Camera $\chi^2$ contingency table.  Cell values contain counts with expected values in parentheses.  Race abbreviations are White (W), Black (B), Native American (NA), Asian American (AA), Hispanic (H), and Other (O).}
\end{table}

%%%%%%%%%%%%%%%%%%%%%%%%%%%%%%%%%%%%%%%%%%%%%%%%%%%
% Discussion
%%%%%%%%%%%%%%%%%%%%%%%%%%%%%%%%%%%%%%%%%%%%%%%%%%%
\section{Discussion}
\noindent \textbf{Summary of Findings:  } The primary goal of this work was to investigate the plausibility of the observed racial distributions of police shooting victims in recent years under various assumptions.  In Sections 3 and 4, we saw that for most races -- White, Black, and Asian, in particular -- the number of shooting victims observed was not at all reasonable to expect based on local population demographics, even when the shooting locations themselves are considered as random.  In Section 5, however, we saw that the observed victim totals are more or less in line with what would be expected when such victims are considered as a random sample from local arrestee populations.  Finally, in Section 6 we observed that the racial distribution of shooting victims appeared to be the same regardless of whether the police officers involved were wearing body cameras.

On the surface, these results appear largely in line with recent findings (e.g.\ Cesario et al.\ (2019) \cite{Cesario2019}).  However, unlike the vast majority of previous work on this topic of which we are aware, the resampling approach employed allows us to consider the issue in greater detail.  Rather than considering only nationwide demographics, the finer-scale approach allows us to incorporate localized information about the populations where these shootings actually took place.  Furthermore, by examining the resampling distributions in their entirety, we can see for example that not only is the proportion of Black victims ``significantly" different from local racial demographics, but that the total number of Black police shooting victims is more than 19 standard deviations larger than expected (see Table 4 in the Appendix).  Perhaps even more surprisingly, incorporating local arrest data shifts this distribution by nearly 17 standard deviations making the observed victim total appear far more in-line.  

While these findings certainly highlight the stark disparities in arrest rates, they are not intended to suggest a racial bias (or lack thereof) on behalf of the police.  On one hand, the reasonably close agreement between observed and expected victim totals when the resampling is done with respect to local arrest demographics might lead one to believe that no such bias exists.  On the other hand, we emphasize the fact that we are utilizing \emph{arrest} rates and not \emph{crime} rates.  Thus, if certain races receive increased attention from law enforcement, then this could, at least in part, potentially explain both higher arrest rates as well as higher proportions of shooting victims.  Stated differently, these findings would support the notion that whatever biases may exist on the arrest level also carry through to the level of fatal shootings. An enormous amount of work has attempted to examine the relationship between race, crime rates, and arrest rates; see \cite{Alessio2003,Smith1984,Beckett2006} for just a few examples. \\

\noindent \textbf{Public Policy Implications:  } Perhaps the most obvious and pressing concern arising in relation to this analysis is the lack of a comprehensive national database on police use of deadly force.  As a result, studies at the national level investigating situational factors most likely to lead to deadly use of force, the kinds of officers most likely to resort to deadly force, and characteristics of individuals most like to have such force used against them are nearly impossible.  Indeed, while some such studies have been carried out at the local level (e.g. \cite{Ridgeway2016}), the biases and woeful underreporting present in current national data largely precludes most larger-scale studies. Here we can only join the chorus of previous researchers (e.g.\ \cite{Kobler1975,Fyfe2002,Klinger2012,Nix2017,Klinger2017,Williams2019,Cesario2019,Klinger2016,White2016} to name just a few) in calling for such data to be collected and made public.  \\

\noindent \textbf{Shortcomings \& Potential Alternative Approaches:  } The WP dataset we rely on contains information on fatal police shooting incidents but not on other forms of fatal police encounters nor on police shootings in which the victim survived.  It is impossible to know or even speculate as to whether the results observed here would extend to this larger set of encounters.  Of minor concern is the slight disagreement between county information contained in the different datasets.  Though every effort was made to correct for spelling, capitalization, and other minor grammatical disparities, some county information remained missing and thus had to be removed or otherwise imputed in the other datasets  (see the Appendix and included \verb!R! file for a complete accounting).  A more significant concern is the differing racial categories across the WP, DEM, and ARREST datasets.  As described throughout, whenever necessary, race categories were imputed according to the U.S. Census Bureau (DEM) data and the relative proportions compared to ensure relatively close agreement.  Nonetheless, because of this disagreement, it is impossible to know to what extent individuals may be labeled differently in the different datasets.  For example, if an individual is White and Hispanic, presumably all of this information would be contained in the DEM dataset but it is unclear and potentially arbitrary as to whether that individual would be categorized as White (W) or Hispanic (H) in the WP dataset as only a single race is provided in each instance.

Also note that in considering the county locations as a random sample, we employed a resampling procedure in which locations were selected according to police officer employment rates.  In doing so, we are, to a degree, making the implicit assumption that the more police officers employed in a given county, the greater the likelihood of a fatal shooting.  Though we felt that this was the most appropriate manner in which to select locations, one could make a reasonable argument that county locations could have instead been resampled according to other characteristics like local arrest rates, violent crime rates, or density of 911 calls for service.  Interestingly, officer employment and arrest totals are positively correlated, though to a lesser degree than one might think at just 0.55.  The same remains true in examining the correlations between officer employment and arrest rates with respect to the individual races included in the ARREST dataset:  W (0.55), B (0.46), NA (0.09), A (0.59).  The code made available with this work may serve as a helpful starting place for researchers wishing to investigate such alternative setups in the future. 

As a final note, it's worth stressing that the preceding sections make no claims or statements involving calculations of the form
\[
P(\text{Suspect is Fatally Shot by Police} \, | \, \text{Suspect Belongs to Race} \,\, r).
\]

That is, we do not investigate statements such as ``members of race $r_i$ are $x$ times more/less likely to be shot by police than members of race $r_j$."  While statements of this sort are quite commonly made in popular media outlets, it's rarely if ever clear how such calculations are made and furthermore, it's not clear that such statements could be reliably made given data of the kind utilized here.  Presumably, the intended claim being made with such statements is that given two races -- $r_i$ and $r_j$ -- and many similar interactions between police and members of these different races, police are more likely to escalate the situation to the point of deploying potentially lethal force if the suspect is of race $r_i$.  Thus, in order to evaluate such claims, one needs data on many police encounters under a variety of situations and outcomes and across each race.  Given enough information of this form, one could presumably evaluate whether the probability of police escalation depends on victim race after taking into account other relevant situational information.  For an alternative analysis that takes this kind of approach, we refer the reader to interesting recent work by Fryer (2016) \cite{Fryer2016}.

%%%%%%%%%%%%%%%%%%%%%%%%%%%%%%%%%%%%%%%%%%%%%%%%%%%%%%%%%%%%
% Acknowledgements
%%%%%%%%%%%%%%%%%%%%%%%%%%%%%%%%%%%%%%%%%%%%%%%%%%%%%%%%%%%%
\section*{Acknowledgements }
A very sincere thank-you to the Washington Post for compiling and providing open access to the dataset of fatal police shootings.  This project was motivated in large part by the availability of such data and likely could not have been completed without such access.  Thank you as well to numerous colleagues and students for providing helpful feedback on early versions of this work.

%%%%%%%%%%%%%%%%%%%%%%%%%%%%%%%%%%%%%%%%%%%%%%%%%%%%%%%%%%%%
% Bibliography
%%%%%%%%%%%%%%%%%%%%%%%%%%%%%%%%%%%%%%%%%%%%%%%%%%%%%%%%%%%%

%%%%%%%%%%%%%%%%%%%%%%%%%%%%%%%%%%%%%%%%%%%%%%%%%%%%%%%%%%%%
% Appendix
%%%%%%%%%%%%%%%%%%%%%%%%%%%%%%%%%%%%%%%%%%%%%%%%%%%%%%%%%%%%
\newpage
\section*{Appendix}
\begin{appendix}

\section{Glossary}
\label{app:A}
{\bf Datasets:}
\begin{itemize}
\item {\bf WP:  }  Dataset containing information on fatal police shootings in the United States between January 2015 and July 2016. 
\item {\bf DEM:  }  Dataset containing information on county-level racial demographics
\item {\bf LEE:  }  Dataset containing information on county-level law enforcement employment totals. 
\item {\bf ARREST:  }  Dataset containing information on county-level crime-related racial demographics
\item {\bf CODES:  }  Dataset containing information on state, county, and parish names along with Uniform Crime Reports (UCR) and Federal Information Processing Standards (FIPS) numeric codes \\
\end{itemize}

\noindent {\bf Racial and Ethnic Demographics:}
\begin{itemize}
\item {\bf W:  } White 
\item {\bf B:   } Black
\item {\bf NA: }  Native American (American Indian) or Alaskan Native
\item {\bf A:   } Asian
\item {\bf NH:   } Native Hawaiian and Other Pacific Islander
\item {\bf HL:   } Hispanic or Latino
\item {\bf T:   } Two or More
\item {\bf O:   } Other
\end{itemize}

\section{Data Access \& Disclosures}
\label{app:B}

We include here a more detailed discussion of modifications made to the data in order to perform the analyses, as well as instructions for how the data may be accessed.  Further details and information can be found in the accompanying \verb!R! file. \\

\noindent {\bf WP:  }  The Washington Post dataset containing information on fatal police shootings in the United States is publicly available via GitHub\footnote{\url{https://github.com/washingtonpost/data-police-shootings}}.  Along with the file containing the data, a readme and data dictionary that discuss the collection methods and variable definitions are also available.  As noted in the main text, the version of this data utilized here was accessed on July 12, 2016 and the most recent shooting recorded at that time was said to have occurred on July 11, 2016.  It is not clear whether modifications are made to prior shootings as the data is updated and thus the current version of the dataset truncated at July 11, 2016 may not match that utilized here exactly.  Except as noted, the WP dataset was kept as original as possible and only small spelling and compatibility changes were made (e.g.\ the town name of ``Ca\~{n}on City'' was changed to ``Canon City'' in order to prevent issues with non-recognizable characters in \verb!R!.)  For various portions of the analysis, some shooting incidents were removed in order to bring the information into agreement with the information available in the other datasets.  Other issues worth noting:
	\begin{itemize}
	\item One recent shooting (ID 1696) in the dataset could not be verified based on the location information provided.  After some searching, it is believed that this corresponds to the shooting that took place in Rush Springs, OK (Grady County) on July 6, 2016.  This information was added.
	\item One shooting victim (ID 1541) was a fugitive.  The shooting occurred in Shawnee National Forest and the county information was recorded as that of the town in which the address is listed (Harrisburg, IL).  
	\item As noted in the main text, a total of 1427 shooting incidents in the WP dataset were utilized in the resampling procedures based on local population demographics and 1249 utilized in the resampling procedures based on local arrest demographics.  The following data descriptions provide more information on which shooting incidents were removed in each analysis and why.
	\item In the event that a city (original location in the WP dataset) crossed into multiple counties, a single county was selected at random for resampling purposes according to population totals (counties with larger populations more likely to be selected).
	\end{itemize}

\noindent {\bf DEM:  }  The county-level racial demography dataset is publicly available and downloadable in csv format (\texttt{cc-est2016-alldata.csv}) through the U.S. Census Bureau website\footnote{\url{https://www.census.gov/data/datasets/2016/demo/popest/counties-detail.html}}.  Subsetting this dataset by YEAR=1 and AGEGRP=0 will produce the 2010 Census data utilized in the above procedures.  Subsetting instead by YEAR=9 will produce the 2016 projections based on the most recent (2010) census.  Summing the male and female totals for each race (columns 11--22) will produce the total population by race and the sum of these should then match the total population (column 8).  Information on the local Hispanic or Latino population is contained in columns 59--70.  As noted in the main text, only one incident in the WP dataset did not appear in the DEM dataset; this was WP ID number 686 which occurred in  Las Cruces, NM in Do\~{n}a Ana County.  Code that takes in the raw dataset and produces the relevant versions used in the above analyses is provided in the accompanying \verb!R! file. \\

\noindent {\bf LEE:  }  The county-level law enforcement employment dataset collected by the FBI through the 2011 Uniform Crime Reporting (UCR) Program is publicly available and downloadable in Microsoft Excel format (\texttt{Table\_80\_Full-time\_Law\_Enforcement\_ \\ Employees\_by\_State\_by\_Metropolitan\_and\_Nonmetropolitan\_Counties\_2011.xls}) through the FBI UCR website\footnote{\url{https://ucr.fbi.gov/crime-in-the-u.s/2011/crime-in-the-u.s.-2011/tables/table-80/view}}.  The raw data has the county information organized by whether the county is considered `Metropolitan.'  This information was moved to a new column which could then be treated as a binary indicator variable (though we note that the `Metropolitan' distinction was not used in any of the analyses).  The original Microsoft Excel file also had a single cell with the state name merged across all other cells corresponding to counties within that state; this was undone and the state name was added to each individual cell.  Note that because of the original non-standard formatting of this dataset, these changes were made outside of the included \verb!R! file.  Other changes to the original data include:
		\begin{itemize}
		\item In Louisiana, the local areas are referred to as parishes instead of counties. The word `Parish' frequently followed area names and was removed to match with other datasets.  In some other instances, the word `County' appeared and was also removed.
		\item Phrases such as ``County Police department" were frequently included.  These were removed in order to match with other datasets.
		\item In the event that multiple entries existed for the same county those totals were summed and the total placed in the row corresponding to county name.
		\item Shannon SD and Do\~{n}a Ana NM were removed because corresponding data was not available in the DEM and WP datasets, respectively. 
		\item Other spelling and grammatical changes include:  ``Augusta-Richmond'' changed to ``Richmond'', ``Hartsville-Trousdale'' changed to ``Trousdale''.  All instances of ``De Kalb'', ``Du Page'' , ``La Salle'', ``La Porte'',  were made into one word.  Various capitalization changes (``Lac Qui Parle'' to ``Lac qui Parle'', ``Lamoure'' to ``LaMoure'', ``Dewitt'' to ``DeWitt'') and spelling changes (``Dillion'' to ``Dillon'' and ``Poweshick'' to ``Poweshiek'' and ``Assymption'' to ``Assumption''). \\
	\end{itemize}

\noindent {\bf ARREST:  }  The county-level arrest data by age, sex, and race is publicly available and downloadable through the Institute for Social Research (ICPSR) at the University of Michigan\footnote{\url{http://www.icpsr.umich.edu/icpsrweb/ICPSR/studies/36115?q=36115}}.  The specific dataset utilized was ICPSR 36115 (\texttt{36115-0001-Data.rda}) and is available to download in a number of convenient formats for use in \verb!R!, SAS, Stata, and SPSS.  The original dataset was subsetted to include only state, county, race, and offense variables.  Counties with missing data were removed.  County identification is given by numeric FBI UCR county code instead of name.  Other issues worth noting: 
	\begin{itemize}
	\item No information is provided from Alaska, Florida, or Washington D.C.  Shooting incidents occurring in these locations were removed as part of the analyses involving arrest rates.
	\item The only arrest information provided from Alabama is from Jefferson County; Cook and Winnebago are the only counties reporting from Illinois.  Shooting incidents occurring elsewhere in these states were removed from the WP dataset in conducting the resampling procedures based on local arrest demographics.  
	\item Other counties appearing in the WP dataset but failing to appear in the ARREST data include, from Virginia: Norfolk, Newport News, Virginia Beach, Suffolk, Hopewell, and Chesapeake; from New York: Queens and Bronx; from Indiana: Jefferson and Orange; from elsewhere, Todd, SD, Davidson, TN, and Broomfield, CO.
	\item Similarly to previous datasets, a variety of spelling and grammatical changes were made in order to bring the datasets into agreement (e.g.\ ``De Kalb" was changed to one word; ``Lagrange" was changed to LaGrange).  Full details can be found in the accompanying \verb!R! code.
	\end{itemize}

\noindent {\bf CODES:  }  The dataset containing both the the FIPS and UCR county codes from the FBI is publicly accessible through the Institute for Social Research (ICPSR) at the University of Michigan\footnote{\url{http://www.icpsr.umich.edu/icpsrweb/NACJD/studies/2565}}.  The specific dataset utilized was the ICPSR 2565 ASCII file which was split into columns according to the ICPSR 2565 Codebook.  As noted in the main text, this dataset was used to match information between the ARREST data which contains only UCR county codes and the WP, DEM, and LEE datasets which contain only the county or parish names.

\newpage
\section{Additional Figures and Tables}
\label{app:C}

\begin{figure}[h]
  \centering
  \includegraphics[scale=0.6]{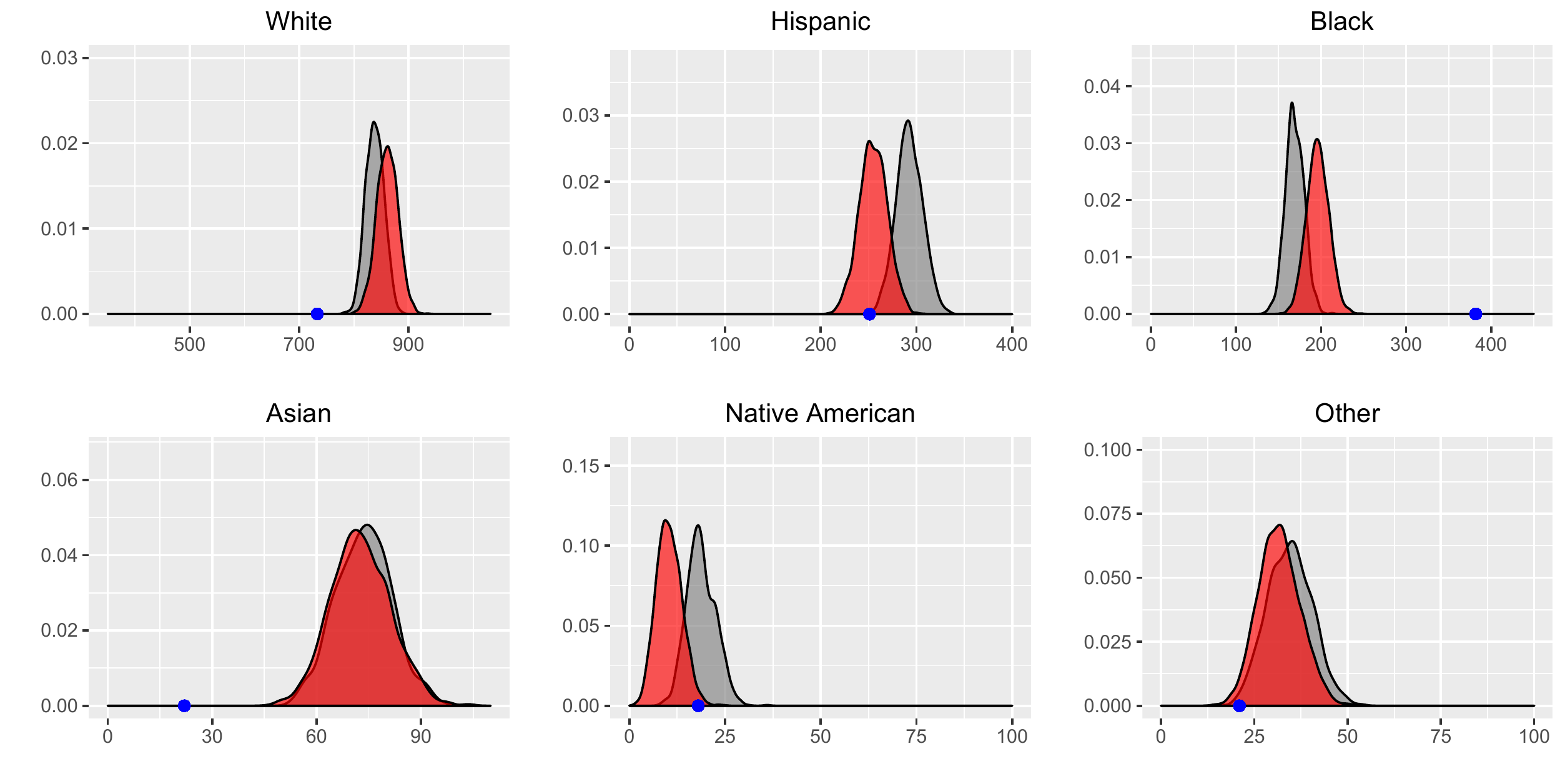}
  \vspace{-7mm}
  \caption{\label{fig:density2016} Kernel density estimates for victim totals by race assuming fixed shooting locations (gray) or random locations (red) according to the 2016 projected racial demographics.  Blue points along each horizontal axis correspond to the true (observed) victim totals from that race in the WP dataset.  Here we see close agreement with Figure \ref{fig:density1} in Section 4.}
\end{figure}

\begin{figure}[h]
  \centering
  \includegraphics[scale=0.6]{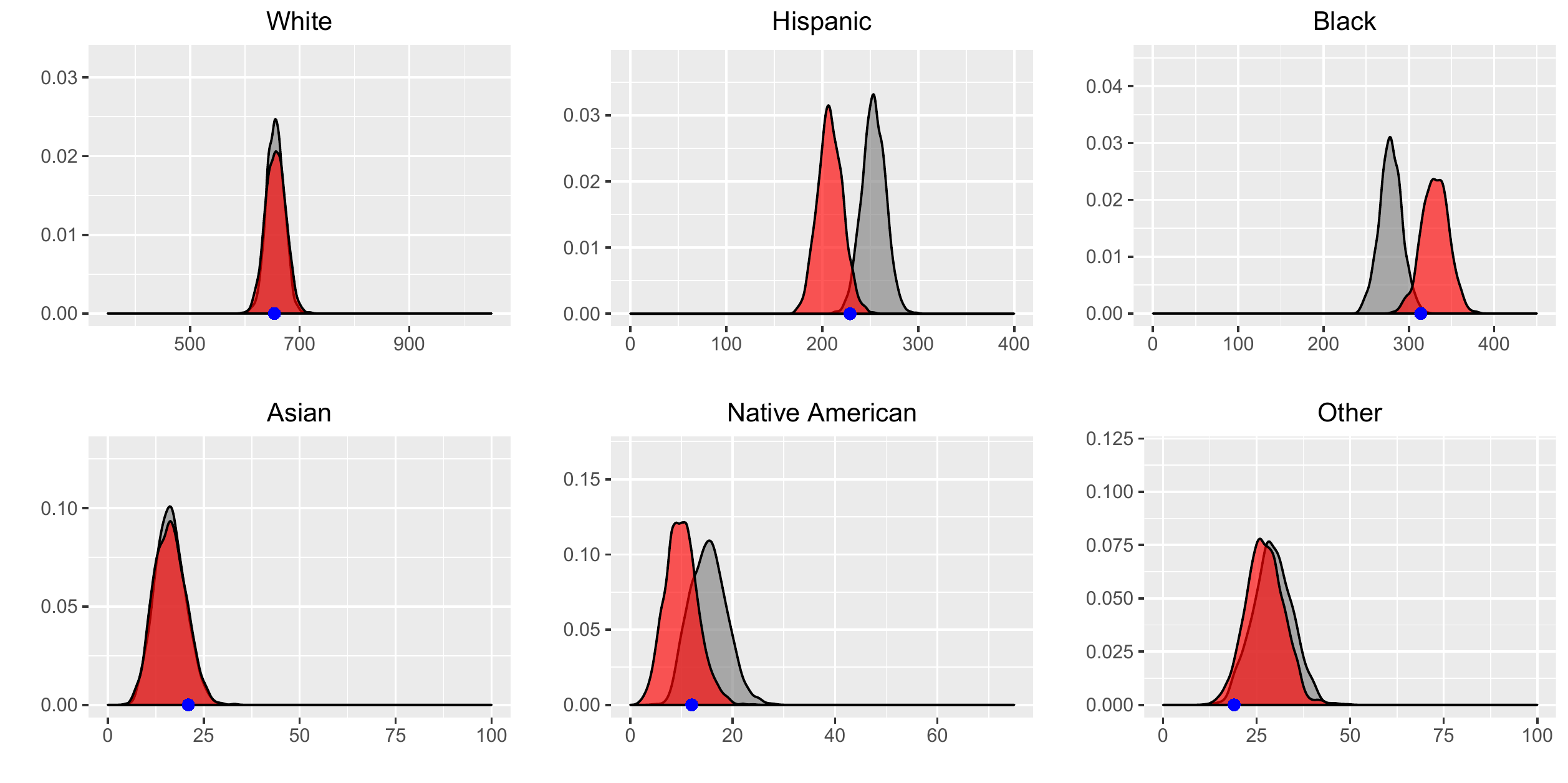}
  \vspace{-7mm}
  \caption{\label{fig:density4} Kernel density estimates for victim totals by race assuming fixed shooting locations (gray) or random locations (red).  Blue points along each horizontal axis correspond to the observed victim totals in the WP dataset.  Victim races selected according to local arrest demographics with races missing arrest totals (Hispanic and Other) sampled according to the 2016 projections.  Here we see close agreement with Figure \ref{fig:density3} in Section 5.}
\end{figure}

\begin{table}[h]
\centering
\begin{tabular}{|l|c|c|c|c|c|}
\hline
\textbf{}                                                                                                   & \textbf{}   & \multicolumn{2}{c|}{\textbf{Fixed Locations}}                                        & \multicolumn{2}{c|}{\textbf{Random Locations}}                                       \\ \hline
\textbf{}                                                                                                   & \textbf{}   & \multicolumn{1}{c|}{\textbf{2010 Census}} & \multicolumn{1}{c|}{\textbf{2016 Proj.}} & \multicolumn{1}{c|}{\textbf{2010 Census}} & \multicolumn{1}{c|}{\textbf{2016 Proj.}} \\ \hline
\multirow{6}{*}{\textbf{\begin{tabular}[c]{@{}l@{}}Population \\ Demographics \end{tabular}}} & \textbf{W}  & 8.2                                   & 6.4                                    & 8.8                                     & 6.6                                    \\ \cline{2-6} 
                                                                                                            & \textbf{H}  & 1.8                                     & 3.0                                    & 0.9                                     & 0.3                                    \\ \cline{2-6} 
                                                                                                            & \textbf{B}  & 19.2                                    & 19.6                                   & 14.7                                    & 14.3                                   \\ \cline{2-6} 
                                                                                                            & \textbf{A}  & 5.5                                     & 6.4                                    & 5.4                                     & 5.9                                    \\ \cline{2-6} 
                                                                                                            & \textbf{NA} & 0.2                                     & 0.2                                    & 2.4                                     & 2.3                                    \\ \cline{2-6} 
                                                                                                            & \textbf{O}  & 1.7                                     & 2.3                                    & 1.3                                     & 1.9                                    \\ \hline
\multirow{6}{*}{\textbf{\begin{tabular}[c]{@{}l@{}} Arrest \\ Demographics \end{tabular}}}     & \textbf{W}  & 1.2                                     & 0.1                                    & 1.0                                     & 0.1                                    \\ \cline{2-6} 
                                                                                                            & \textbf{H}  & 0.6                                     & 2.0                                    & 2.8                                     & 1.6                                    \\ \cline{2-6} 
                                                                                                            & \textbf{B}  & 2.5                                     & 2.7                                    & 1.2                                     & 1.1                                    \\ \cline{2-6} 
                                                                                                            & \textbf{A}  & 1.2                                     & 1.2                                    & 1.2                                     & 1.2                                    \\ \cline{2-6} 
                                                                                                            & \textbf{NA} & 1.0                                     & 0.9                                    & 0.5                                     & 0.6                                    \\ \cline{2-6} 
                                                                                                            & \textbf{O}  & 1.415                                     & 2.0                                    & 1.0                                     & 1.6                                    \\ \hline
\end{tabular}
\caption{\label{tab:SDdistances} Number of standard deviations the observed victim totals lie from the expected victim totals for each resampled density.}
\end{table}

\clearpage
\section{Randomization Test for Body Camera Effect}
\label{app:D}

As noted in the main text, here we are interested in whether the distribution of victims across race is the same when police-worn body cameras are vs.\ are not in use and thus a standard $\chi^2$ testing approach is natural.  While the same approach could have, in theory, been used in the previous sections, we instead employed randomization tests as an intuitive and clean workaround to the high dimensional issues that would have been introduced.  In the interest of completeness, we now carry out a randomization-style test to examine body camera effects as well.  

From Table \ref{tab:chisq}, we see that there are 1296 fatal shooting incidents in which no body camera was present and only 132 in which a police-worn body camera was present.  The higher number of no-body-camera incidents means that the resulting observed victim proportions in these cases should be more accurate and thus we treat this as our reference distribution.  As in the earlier analyses, we sample a race at random weighted according to these proportions and repeat the process 132 times in order to simulate  a single count distribution under the null hypothesis that the distributions are the same regardless of body-camera presence.  The process is repeated 1000 times in order to generate null distributions of victim totals for each race and we then compare the observed counts to these distributions.  

\begin{figure}
  \centering
  \includegraphics[scale=0.6]{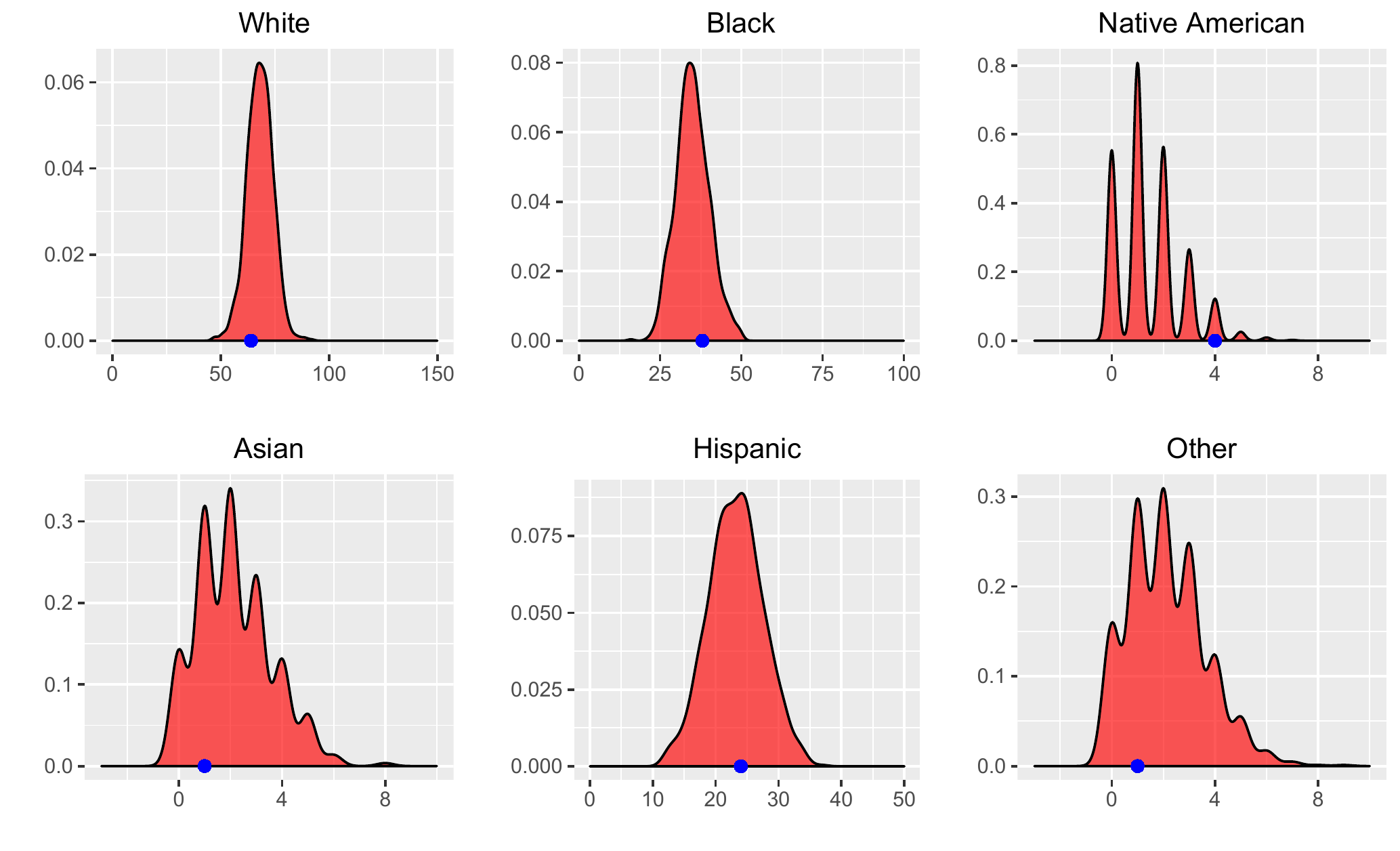}
  \vspace{-7mm}
  \caption{\label{fig:BodyCamDensity} Kernel density estimates for victim totals by race assuming the distribution is the same as in instances where no body camera is present.  Blue points along each horizontal axis correspond to the observed victim totals in the WP dataset.}
\end{figure}

The resulting distributions are shown in Figure \ref{fig:BodyCamDensity} with p-values for each race given in Table \ref{tab:BodyCamRand}.  The randomization tests suggest a significant effect at the $0.05$ level only for Native Americans; no significant results remain after a Bonferroni correction.  We stress however the sensitivity of such results for races with low victim totals as even the density estimates take on a more discrete appearance.  For example, had there been 3 Native American victims rather than the 4 observed, this result would no longer be deemed significant at the 0.05 level.

\begin{table}
\centering
\begin{tabular}{|c|c|c|c|c|c|c|}
\hline
                 & W    & B     & NA    & A    & H     & O     \\ \hline
\textit{p}-value & 0.55 & 0.472 & 0.032 & 0.74 & 0.772 & 0.752 \\ \hline
\end{tabular}
\caption{\label{tab:BodyCamRand} P-values computed for each race resulting from the randomization tests evaluating the effect of police-worn body cameras.  Only the result for Native Americans is significant at the $0.05$ level; no significant results remain after applying a Bonferroni correction.}
\end{table}

\end{appendix}

\end{document}